\documentstyle[12pt,epsfig]{article}
\newcommand{\simlt}  {\raisebox{-.6ex}{$\stackrel{\textstyle <}{\sim}$}}

\begin{document}
\begin{flushright}
hep-ph/0203209 \\
RAL-TR-2002-008 \\
21 Mar 2002 \\
\end{flushright}
\begin{center}
{\Large
Symmetries and Generalisations of Tri-Bimaximal 
Neutrino~Mixing.
}
\end{center}
\vspace{1mm}
\begin{center}
{P. F. Harrison\\
Physics Department, Queen Mary and Westfield College\\
Mile End Rd. London E1 4NS. UK \footnotemark[1]}
\end{center}
\begin{center}
{and}
\end{center}
\begin{center}
{W. G. Scott\\
Rutherford Appleton Laboratory\\
Chilton, Didcot, Oxon OX11 0QX. UK \footnotemark[2]}
\end{center}
\vspace{1mm}
\begin{abstract}
\baselineskip 0.6cm
\noindent
Tri-bimaximal mixing
is a specific lepton mixing ansatz,
which has been shown to account
very successfully
for the established
neutrino oscillation data. 
Working in a particular basis
(the `circulant basis'),
we identify three 
independent symmetries 
of tri-bimaximal mixing,
which we exploit
to set the tri-bimaximal hypothesis in context,
alongside some simple,
phenomenologically interesting
$CP$-conserving and $CP$-violating 
generalisations. 
\end{abstract}
\begin{center}
\end{center}
\footnotetext[1]{E-mail:p.f.harrison@qmul.ac.uk}
\footnotetext[2]{E-mail:w.g.scott@rl.ac.uk}
\newpage
\baselineskip 0.6cm

\noindent {\bf 1. Introduction} \\

\noindent
Tri-bimaximal mixing \cite{hps5}
is a very successful lepton mixing ansatz,
which has already attracted a degree of attention
in the literature \cite{xing}.
In the standard parametrisation \cite{pdg},
tri-bimaximal mixing may be specified by:
$\theta_{12}=\sin^{-1}(1/\sqrt{3})$,
$\theta_{23}=-\pi/4$ and $\theta_{13}=0$,
with no $CP$-violating phase.
Tri-bimaximal mixing
builds squarely on all of the most promising
phenomenological ideas which have preceded it
\cite{trix}
and readily accounts for all
of the best-established 
neutrino oscillation results to date
\cite{sols} \cite{amos} \cite{reac}.

Despite these sucesses 
we have no reason to suppose
that tri-bimaximal mixing 
will prove to be exactly right
in every detail,
and we seek therefore to generalise
the original tri-bimaximal hypothesis,
so as to parametrise possible deviations
in simple and meaningful ways,
which we hope will be useful
in developing experimental tests.
We begin by reviewing the symmetries
inherent in the tri-bimaximal scheme,
which will lead us to 
identify generic features
which will form the basis 
of our generalisations. \\

\noindent {\bf 2.  
Tri-Bimaximal Mixing and the Circulant Basis } \\

\noindent
Symmeties are usually thought
to be best studied at the level
of the mass-matrices,
which are naturally 
referred to a `weak' basis
(ie. a basis which leaves 
the charged-current
weak-interaction 
diagonal and universal).
Furthermore,
by restricting 
consideration to left-handed fields only,
we may take our mass-matrices (squared) 
to be hermitian.
Following Ref.~\cite{hps5},
we will work in a particular weak basis 
in which the mass-matrix for the charged leptons
takes the familiar $3 \times 3$ 
`circulant' form \cite{hs1}:
\begin{equation}
M_l^2 =
\left(\matrix{
a & b & b^{*} \cr
b^{*} & a & b \cr
b & b^{*} & a \cr
} \right)
\label{ml2}
\end{equation}
where the constants $a$, $b$ and $b^*$ 
encode the lepton masses as follows:
\begin{eqnarray}
a=\frac{m_e^2}{3} +\frac{m_{\mu}^2}{3}+\frac{m_{\tau}^2}{3} \nonumber \\
b=\frac{m_e^2}{3} +\frac{m_{\mu}^2 \omega}{3}
                  +\frac{m_{\tau}^2 \bar{\omega}}{3} \label{abemt} \\
b^*=\frac{m_e^2}{3} +\frac{m_{\mu}^2 \bar{\omega}}{3}
                  +\frac{m_{\tau}^2 \omega}{3} \nonumber
\end{eqnarray}
($\omega=\exp(i2\pi/3)$ and $\bar{\omega}=\exp(-i2 \pi/3)$
are complex cube roots of unity).

In the above basis 
(the `circulant basis')
the charged-lepton mass-matrix (Eq.~\ref{ml2})
is clearly invariant under
cyclic permutations
of the three generation indices.
Note, however, that invariance 
under {\em odd} permuations
of the generation indices
(ie.\ generation interchange)
would require that {\em odd} permutations
are performed simultaneously 
with a complex conjugation 
(see Section~3 below).

Thus far,
we have simply chosen a basis, 
and it is 
the form of the neutrino mass matrix
in this basis
which determines 
the observable mixing.
Since circulant matrices 
of identical order always commute,
we cannot take the neutrino mass-matrix
to be also a $3 \times 3$ circulant
and obtain non-trivial mixing.
We were led therefore 
to postulate \cite{hps5}
a neutrino mass-matrix
which is invariant
under cyclic permutations 
of only {\em two} 
out of the three generations,
ie.\ a $2 \times 2$ circulant
with one generation
(which was taken
to be generation 2
in this basis)
isolated in the mass-matrix 
by four `texture zeroes' \cite{ross},
yielding the effective block-diagonal form:
\begin{equation}
\hspace{2.0cm}
M_{\nu}^2 =
\left(\matrix{
x & 0 & y \cr
0 & z & 0 \cr
y & 0 & x \cr
} \right). \hspace{2.0cm} \label{mn2}
\end{equation}
In Eq.~\ref{mn2}
the constants
$x$, $y$ and $z$ ($y$ negative) 
encode the neutrino masses as follows:
\begin{eqnarray}
x=\frac{m_1^2}{2} +\frac{m_{3}^2}{2} \nonumber \\
y=\frac{m_1^2}{2} -\frac{m_{3}^2}{2} \label{123xy} \\
z=m_2^2 \nonumber
\end{eqnarray}
Note that the neutrino mass matrix 
Eq.~\ref{mn2} 
is real and symmetric
(as well as being,
by construction, 
a $2 \times 2$
circulant in the $1-3$ index subset).
With $y$ real, 
the neutrino mass-matrix Eq.~\ref{mn2}
is invariant under
the {\em odd} permutation 
corresponding to the generation
interchange $1 \leftrightarrow 3$,
performed with or without
a complex conjugation 
(cf.\ Eq.~\ref{ml2} above).

The charged-lepton mass-matrix 
$M_l^2$ (Eq.~\ref{ml2})
and the neutrino mass-matrix 
$M_{\nu}^2$ (Eq,~\ref{mn2}) 
are diagonalised by threefold maximal and twofold maximal
unitary matrices $U_l$ and $U_{\nu}$, respectively \cite{hps5}.
The MNS matrix \cite{mns} is then given by 
$U_l^{\dagger}U_{\nu}=U$ :
\begin{eqnarray}
    \matrix{  \hspace{0.2cm} \nu_1 \hspace{0.2cm}
               & \hspace{0.2cm} \nu_2 \hspace{0.1cm}
               & \hspace{0.2cm} \nu_3  \hspace{0.2cm} }
                                      \hspace{0.9cm}
\hspace{1,6cm}
    \matrix{  \hspace{0.2cm} \nu_1 \hspace{0.2cm}
               & \hspace{0.2cm} \nu_2 \hspace{0.2cm}
               & \hspace{0.2cm} \nu_3  \hspace{0.2cm} }
                                      \hspace{0.9cm} \nonumber \\
\matrix{ e \hspace{0.2cm} \cr
         \mu \hspace{0.2cm} \cr
         \tau \hspace{0.2cm} }
\left(\matrix{
\frac{1}{\sqrt{3}}
                & \frac{1}{\sqrt{3}} & \frac{1}{\sqrt{3}} \cr
\frac{\bar{\omega}}{\sqrt{3}}
                & \frac{1}{\sqrt{3}}
                    & \frac{\omega}{\sqrt{3}} \cr
\frac{\omega}{\sqrt{3}}
             & \frac{1}{\sqrt{3}}
                    & \frac{\bar{\omega}}{\sqrt{3}} \cr
} \right)
\hspace{0.2cm}
\left(\matrix{
\sqrt{\frac{1}{2}} & 0 & \sqrt{\frac{1}{2}} \cr
 0 & 1 & 0 \cr
\sqrt{\frac{1}{2}} & 0  & -\sqrt{\frac{1}{2}} \cr
} \right)
\hspace{0.35cm}
=
\hspace{0.35cm}
\matrix{ e \hspace{0.2cm} \cr
         \mu \hspace{0.2cm} \cr
         \tau \hspace{0.2cm} }
\left(\matrix{
\sqrt{\frac{2}{3}} & \sqrt{\frac{1}{3}} &  0 \cr
-\sqrt{\frac{1}{6}} & \sqrt{\frac{1}{3}} & -\frac{i}{\sqrt{2}} \cr
-\sqrt{\frac{1}{6}}& \sqrt{\frac{1}{3}}   & \frac{i}{\sqrt{2}} \cr
} \right) \hspace{0.80cm}
\label{decomp}
\end{eqnarray}
which is 
tri-bimaximal mixing
in a particular phase convention. 
Note that the 
tri-bimaximal mixing matrix
(Eq.~\ref{decomp} - RHS)
has two rows (row 2 and row 3)
which are complex conjugates of each other
(so that corresponding elements are equal in modulus).
This is readily traced to the fact
that the matrix $U_l^{\dagger}$ 
(Eq.~\ref{decomp} - LHS) 
has likewise two rows (row 2 and row 3)
complex-conjugate,
while the matrix $U_{\nu}$ is real.
It will prove useful to observe
(see Section 3)
that the matrix $U_l^{\dagger}$ (Eq.~\ref{decomp} - LHS)
has also two {\em columns}
(columns 1 and 3) 
complex conjugate. \\

\noindent {\bf 3.
CP-Conservation and Tri-Phi-Maximal Mixing} \\

\noindent
In the circulant basis
(see Section 2, above) 
a number of generic features
of tri-bimaximal mixing (Eq.~\ref{decomp})
manifest themselves very simply
in the form
of the neutrino mass matrix (Eq.~\ref{mn2}). 
Perhaps the most significant feature
of tri-bimaximal mixing is the predicted 
absence of $CP$-violation 
in neutrino oscillations.
If $CP$ is conserved,
the MNS matrix is orthogonal
(or may be taken\footnote{
The phases of the charged-lepton
mass-eigenstates are entirely unphysical
and may be re-defined at will.
Also,
the phases of the neutrino mass-eigenstates
have no influence on the form of the
neutrino mass-matrix (squared) $M_{\nu}^2$ 
as defined here, 
ie.\ no influence on Eq.~\ref{jcp0}. }  
to be orthogonal).
As remarked in Section~2,
the unitary matrix 
$U_l^{\dagger}$ (Eq.~\ref{decomp} - LHS)
has two columns (columns 1 and 3)
complex conjugates of each other.
If the MNS matrix is orthogonal
(ie. $U=O$),
we have $U_{\nu}=U_l O$
which then gives $U_{\nu}$ with two {\em rows}
(rows 1 and 3) 
complex conjugates of each other.
Since 
$M_{\nu}^2 = U_{\nu}{\rm diag}(m_1^2, \; 
m_2^2, \; m_3^2) U_{\nu}^{\dagger}$
this symmetry
must then be manifest
in the neutrino mass-matrix
so that, in particular,
a sufficiently general form
for the neutrino mass matrix
in the circulant basis,
yielding no CP-violation
in neutrino oscillations,
may be written\footnote{ 
It should perhaps be said that 
the similarity of 
Eq.~\ref{jcp0} of the present paper
to Eq.~31 of Ref.~\cite{grim} 
appears to be somewhat accidental.
We remind the reader that 
we are working here in the `circulant basis'
(defined in Section~2, above) 
and not in the lepton flavour basis
as in Ref.~\cite{grim}.}:
\begin{equation}
\hspace{2.0cm}
M_{\nu}^2 =
\left(\matrix{
x & w & y^* \cr
w^* & z & w \cr
y & w^* & x \cr
} \right). \hspace{2.0cm} \label{jcp0}
\end{equation}
The form Eq.~\ref{jcp0}
generalises Eq.~\ref{mn2},
and also mirrors the circulant form Eq.~\ref{ml2}
but with the $2$nd generation distinguished
(when $z=x$ and $y=w$
Eq.~\ref{jcp0} becomes circulant).
The six real parameters 
correspond to the three masses
and three real mixing angles.
The important point 
however is that Eq.~\ref{jcp0}
exhibits the symmetry of Eq.~\ref{ml2}
under the exchange of generations 
$1 \leftrightarrow 3$
performed simutaneously 
with a complex conjugation.
Indeed, in the circulant basis,
it is the invariance
of {\em all} the leptonic terms
under this combined operation
that ensures $J_{CP}=0$ 
(since Im $\det [M_l^2,M_{\nu}^2]$ changes sign).

Starting from Eq.~\ref{jcp0}, 
if we take $w$ and $y$ real,
we immediately recover
Altarelli-Feruglio mixing \cite{altf}
with $\tan 2\theta = 2\sqrt{2}(x+y-z-w)/(x+y-z+8w)$.
If we consider only the 
combination $w^2y$ real
(so that $w$ and $y$ 
have correlated phase, with
$w=|w|\exp(-i\phi)$, $y=-|y|\exp(i2\phi)$ complex)
we obtain a simple two parameter
generalisation of Altarelli-Feruglio mixing
with $J_{CP}=0$, but $|U_{e3} | \ne 0$ in general.
Taking $|w| \rightarrow 0$
(as required for tri-bimaximal mixing) 
but keeping $y=-|y|\exp(i2\phi)$ 
complex
with a fixed phase-angle $\phi$,
we obtain a block-diagonal, complex
neutrino mass-matrix, generalising 
Eq.~\ref{mn2}, which will lead
to `tri-$\phi$maximal' mixing (below):
\begin{equation}
\hspace{2.0cm}
M_{\nu}^2 =
\left(\matrix{
x & 0 & y^* \cr
0 & z & 0 \cr
y & 0 & x \cr
} \right). \hspace{2.0cm} \label{mnh2}
\end{equation}
The neutrino masses 
are given similarly to Eq.~\ref{123xy}
(but with $y$ replaced by $-|y|$):
\begin{eqnarray}
x=\frac{m_1^2}{2} +\frac{m_{3}^2}{2} \nonumber \\
|y|=\frac{m_3^2}{2} -\frac{m_1^2}{2} \\
z=m_2^2 \nonumber
\end{eqnarray}
and we obtain the simple,
one-parameter,
$CP$-conserving 
generalisation of tri-bimaximal mixing
(depending on the phase-angle $\phi$)
referred to here as `tri-$\phi$maximal' mixing:
\begin{eqnarray}
     \matrix{  \hspace{0.4cm} \nu_1 \hspace{1.7cm}
               & \hspace{0.4cm} \nu_2 \hspace{1.4cm}
               & \hspace{0.4cm} \nu_3  \hspace{0.4cm} }
                                      \hspace{1.4cm} \nonumber \\
\hspace{1.0cm}U \hspace{0.3cm} = \hspace{0.3cm} 
\matrix{ e \hspace{0.2cm} \cr
         \mu \hspace{0.2cm} \cr
         \tau \hspace{0.2cm} }
\left( \matrix{ \sqrt{\frac{2}{3}}\cos \phi  &
                      \frac{1}{\sqrt{3}} &
                               \sqrt{\frac{2}{3}} \sin{\phi} \cr
 - \frac{\cos \phi}{\sqrt{6}}  - \frac{\sin \phi}{\sqrt{2}} &
          \frac{1}{\sqrt{3}} &
  \frac{\cos \phi}{\sqrt{2}}  - \frac{\sin \phi}{\sqrt{6}} \cr
      \hspace{2mm} 
  - \frac{\cos \phi}{\sqrt{6}}  + \frac{\sin \phi}{\sqrt{2}}
                                                   \hspace{2mm} &
         \hspace{2mm} 
         \frac{1}{\sqrt{3}} \hspace{2mm} &
 -\frac{\cos \phi}{\sqrt{2}}  - \frac{\sin \phi}{\sqrt{6}} 
                                       \hspace{2mm} \cr } \right).
\hspace{1.0cm}
\label{phb2}
\end{eqnarray}
The phase angle $\phi$ 
must satisfy
$|\sin \phi|$ $\simlt$ $0.2$
to fit the reactor data \cite{reac},
while there is minimal impact
on the fit to the atmospheric data \cite{amos}.  
Tri-$\phi$maximal mixing has $U_{e3} \ne 0$,
but retains the symmetries 
$|U_{e 2}| = |U_{\mu 2}| = |U_{\tau 2}|$  
and $J_{CP}=0$.
Note that in tri-$\phi$maximal mixing
the symmetry of two rows
complex conjugate is sacrificed
(compare Eq.~\ref{phb2}
rows 2 and 3 
with Eq.~\ref{decomp} - RHS,
rows 2 and 3).
Clearly tri-$\phi$maximal mixing
reduces to 
tri-bimaximal mixng 
in the limit $\phi \rightarrow 0$. \\

\noindent {\bf 4. CP-Violation and Tri-Chi-Maximal Mixing} \\

\noindent
In the circulant basis,
it will be enough to require
that the neutrino mass matrix be real
(ie.\ symmetric, 
since our mass matrices are hermitian)
to ensure that two rows of the MNS matrix
have corresponding elements 
which are  
equal in modulus~\footnote{
In general, if the MNS matrix has two rows
with corresponding elements equal in modulus,
then by approriate re-phasings, 
the two rows may always be taken
to be complex-congugate to each other,
with the remaining row taken to be purely real.},
just as in Eq.~1 
(rows 2 and 3 in this case).
Such mixing matrices 
form an already interesting (two-parameter) 
generalisation of tri-bimaximal mixing,
with a form of mu-tau universality.
This time (cf.\ Section~3 above) 
the proof depends
on the unitary matrix $U_l^{\dagger}$ (Eq.~\ref{decomp})
having two {\em rows} 
(row 2 and row 3) complex-conjugate.
The theorem
follows immediately by noting that
a real symmetric matrix
may always be diagonalised 
by an orthogonal matrix 
(ie.\ $U_{\nu}=O$, with $O$ an orthogonal matrix)
so that the resulting MNS matrix
$U \equiv U_l^{\dagger}U_{\nu}=U_l^{\dagger}O$,
is necessarily also of the form
Eq.~\ref{decomp},
with rows 2 and 3 complex-conjugate
and row 1 purely real.
Clearly the resulting mixing matrix
is invariant 
under interchange of row 2 and row 3 
performed simultaneously 
with a complex-conjugation.
This
form of mu-tau univesality 
implies strict mu-tau symmetry 
for $CP$-even observables
(eg.\ disappearance probabilities),
but has 
$CP$-odd observables
(eg.\ asymmetries in appearance probabiliites) 
changing sign.

Of course, 
non-zero $CP$-violation
means sacrificing the symmetry 
(Eq.~\ref{jcp0})
of the neutrino mass-matrix
under 
$1 \leftrightarrow 3$ interchange
together with complex-conjugation.
If we do this
taking the 
neutrino mass-matrix to be real
as discussed above,
and again demanding
effective block-diagonal form,
just as for tri-bimaximal mixing,
we are immediately led
to the neutrino mass-matrix
for `tri-$\chi$maximal' mixing:
\begin{equation}
\hspace{2.0cm}
M_{\nu}^2 =
\left(\matrix{
x & 0 & y \cr
0 & z & 0 \cr
y & 0 & w \cr
} \right) \hspace{2.0cm} \label{mna2}
\end{equation}
The real constants $x$, $y$, $z$ and $w$ 
now encode the neutrino masses and 
one mixing angle $\chi$ as follows:
\begin{eqnarray}
x=\frac{m_1^2+m_3^2}{2}-\frac{m_3^2-m_1^2}{2} \sin 2 \chi \nonumber \\
w=\frac{m_1^2+m_3^2}{2}+\frac{m_3^2-m_1^2}{2} \sin 2 \chi \label{a2123} \\
z=m_2^2 \nonumber
\end{eqnarray}
where $\cot 2 \chi = 2y/(x-w)$,
leading to our second
simple one-parameter generalisation
of tri-bimaximal mixing,
called here `tri-$\chi$maximal' mixing:
\begin{eqnarray}
     \matrix{  \hspace{0.4cm} \nu_1 \hspace{1.7cm}
               & \hspace{0.4cm} \nu_2 \hspace{1.4cm}
               & \hspace{0.4cm} \nu_3  \hspace{0.4cm} }
                                      \hspace{1.4cm} \nonumber \\
\hspace{1.0cm}U \hspace{0.3cm} = \hspace{0.3cm} 
\matrix{ e \hspace{0.2cm} \cr
         \mu \hspace{0.2cm} \cr
         \tau \hspace{0.2cm} }
\left( \matrix{ \sqrt{\frac{2}{3}}\cos \chi  &
                      \frac{1}{\sqrt{3}} &
                               \sqrt{\frac{2}{3}} \sin{\chi} \cr
 - \frac{\cos \chi}{\sqrt{6}}  - i \frac{\sin \chi}{\sqrt{2}} &
          \frac{1}{\sqrt{3}} &
   i \frac{\cos \chi}{\sqrt{2}}  -  \frac{\sin \chi}{\sqrt{6}} \cr
      \hspace{2mm} 
   - \frac{\cos \chi}{\sqrt{6}}  + i \frac{\sin \chi}{\sqrt{2}}
                                                   \hspace{2mm} &
         \hspace{2mm} 
         \frac{1}{\sqrt{3}} \hspace{2mm} &
 - i \frac{\cos \chi}{\sqrt{2}}  -  \frac{\sin \chi}{\sqrt{6}} 
                                       \hspace{2mm} \cr } \right)
\hspace{1.0cm}
\label{pha2}
\end{eqnarray}
Again, we have $|\sin \chi|$ $\simlt$ $0.2$ 
in order to fit the reactor data \cite{reac}.
Tri-$\chi$maximal mixing has 
non-zero $U_{e3}= \sqrt{2/3}\sin \chi$ 
and maximal $CP$-violation 
(for fixed $|U_{e3}|$) with
the Jarlskog invariant \cite{jarl}
given by
$J_{CP} = \sin 2 \chi/(6 \sqrt{3}$).
As expected, Eq.~\ref{pha2}
has rows 2 and 3 complex conjugates
of each other.
Tri-$\chi$maximal mixing Eq.\ref{pha2}
and tri-$\phi$maximal mixing  Eq.~\ref{phb2}
are clearly very closely related
(they are identical
interchanging
$\chi \leftrightarrow \phi$,
apart from the factors of $i$).
Again,
tri-$\chi$maximal mixing (Eq.~\ref{pha2})
reduces to tri-bimaximal mixing 
in the limit $\chi \rightarrow 0$.

Finally,
we note that there is always the possibility
of specialising our mixings
by imposing additional constraints. 
An amusing
specialisation of tri-$\chi$maximal mixing  
would be to require $y=z-(x+w)/2$, which leads immediately to:
$\sin \chi =\sqrt{\Delta m^2_{21}/\Delta m^2_{31}} \sim 0.13$,
certainly consistent with current experimental limits \cite{reac}, 
and holding out the promise 
of observable $CP$-violation 
in future experiments \cite{edge}. \\

\noindent {\bf 5. Discussion} \\

\noindent
We have identified three symmetries
of tri-bimaximal mixing.
In the circulant basis
(where the charged-lepton mass-matrix
takes a simple $3 \times 3$ circulant form)
these symmetries may be (separately) implemented
by taking the neutrino mass-matrix to be 
i)~real,
ii)~invariant under $1 \leftrightarrow 3$ 
   interchange with complex conjugation and
iii)~effective block-diagonal
   in the $1,3$ index subset. 
At the level of the mixing matrix
these symmetries correspond to the
properties
i)~two rows (rows 2 and 3)
   complex conjugate (equal in modulus),
ii)~no CP-violation 
   in neutrino oscillations ($J_{CP}=0$) and
iii)~trimaximal mixing for solar neutrinos, ie.\ 
$|U_{e 2}| = |U_{\mu 2}| = |U_{\tau 2}| =1/\sqrt{3}$.

Together with the Altarelli-Feruglio hypothesis \cite{altf},
tri-$\phi$maximal and tri-$\chi$maximal mixing
form the complete set of natural
one-parameter generalisations
of tri-bimaximal mixing,
defined by dropping any one
(or retaining any two)
of the above three symmetries.
Thus tri-$\phi$maximal mixing (Eq.~\ref{phb2})
retains ii)~$J_{CP}=0$ and 
iii)~$|U_{e 2}| = |U_{\mu 2}| = |U_{\tau 2}| =1/\sqrt{3}$,
but drops i)~two-rows complex-conjugate,
while tri-$\chi$maximal mixing (Eq.~\ref{pha2})
retains i)~two-rows complex-conjugate 
and 
iii)~$|U_{e 2}| = |U_{\mu 2}| = |U_{\tau 2}|=1/\sqrt{3}$,
dropping ii)~$J_{CP}=0$.
Altarelli-Feruglio mixing \cite{altf} 
completes the set, dropping 
iii)~$|U_{e 2}| = |U_{\mu 2}| = |U_{\tau 2}|=1/\sqrt{3}$,
but retaining
i)~two rows complex-conjugate (equal in modulus)
and ii)~$J_{CP}=0$.
Of course, only tri-bimaximal mixing itself
retains all three of the above symmetries
(and is furthermore completely defined by them).

There is no implication here
that the list of one-parameter
generalisations
of tri-bimaximal mixing 
is exhausted.
For example,
we have also considered 
$CP$-conserving and $CP$-violating
analogues of 
Eq.~\ref{phb2} and Eq.~\ref{pha2}
which leave unmodified
the first column
(rather than the second column)
of the tri-bimaximal mixing matrix,
altough we judge these
somewhat less `natural'
than those presented above.
 
Clearly, 
less-constrained ansatze are obtained 
dropping two symmetries simultaneously,
ie.\ retaining only one.
The (three-parameter) 
set of all mixings with $J_{CP}=0$
(see Section~3)
is obviously too unconstrained 
to be useful,
and the two parameter mixing ansatz,
having rows 2 and 3
of the MNS matrix complex-conjugate
(equal in modulus)
has already been discussed in Section~4.
Our final,  
and perhaps our most useful,
two-parameter generalisation 
of tri-bimaximal mixing drops 
i) two rows complex conjugate
and
ii) $J_{CP}=0$,
but retains
iii)~$|U_{e 2}| = |U_{\mu 2}| = |U_{\tau 2}|=1/\sqrt{3}$.
This mixing ansatz
is defined (in the circulant basis) 
simply by a neutrino mass-matrix
with four off-diagonal `texture zeroes' \cite{ross}
(ie.\ by a neutrino mass-matrix 
with effective block-diagonal form,
cf.\ Eq.~\ref{mn2}),
and it interpolates smoothly between 
tri-$\phi$maximal and tri-$\chi$maximal mixing.
Appealing to the unitarity of the MNS matrix,
one might even claim that
the block-diagonal constraint
(ie.\ the presence of the 
texture zeroes, Eq.~\ref{mn2}) 
is enough to {\em explain} the solar data \cite{sols} 
{\em and} to explain near maximal mu-tau mixing
at the atmospheric scale \cite{amos}, 
given $|U_{e3}|$ small from reactors \cite{reac}.

We do not know
which (if any) of the above symmetries
will survive
as experiments become more refined. 
With $CP$-violation
an established feature
of quark mixing \cite{kaon} \cite{baba},
and $CP$-violation in the lepton-sector 
seen universally as 
a crucial goal experimentally \cite{edge},
we are tempted to give
most emphasis here
to our one-parameter $CP$-violating ansatz
`tri-$\chi$maximal' mixing, Eq.~\ref{pha2},
as our most interesting 
and predictive ansatz.
In any case, however,
in practical terms,
tri-$\phi$maximal mixing
and 
tri-$\chi$maximal mixing 
represent
the two extremes that
one has necessarily to consider
experimentally,
and we have also made it
clear how best to interpolate
between them.
Perhaps
the most remarkable thing is that
tri-bimaximal mixing itself
(which adequately represents
current experimental observation)
comprises so many symmetries.

\vspace{5mm}
\noindent {\bf Acknowledgement}

\noindent
This work was supported by the UK
Particle Physics and Astronomy Research Council
(PPARC).

\end{document}